\documentclass[twocolumn,graphics,prl,aps,floats,superscriptaddress,showpacs]{revtex4}
\usepackage{graphicx}

\begin{document}

\title{Heat Conduction in the Vortex State of NbSe$_2$: \\
Evidence for Multi-Band Superconductivity}

\author{Etienne Boaknin}
\author{M.A. Tanatar}
\author{Johnpierre Paglione}
\author{D. Hawthorn}
\author{F. Ronning}
\author{R.W. Hill}
\author{M. Sutherland}
\affiliation{Department of Physics, University of Toronto,
Toronto, Ontario, Canada}
\author{Louis Taillefer}
\affiliation{Department of Physics, University of Toronto,
Toronto, Ontario, Canada} \affiliation{Canadian Institute for
Advanced Research, Toronto, Ontario, Canada}
\author{Jeff Sonier}
\affiliation{Canadian Institute for Advanced Research, Toronto,
Ontario, Canada}
\affiliation{Simon Fraser University, Department
of Physics, Burnaby, Canada}
\author{S.M. Hayden}
\affiliation{H. H. Wills Physics Laboratory, University of
Bristol, Bristol BS8 1TL, United Kingdom}
\author{J.W. Brill}
\affiliation{Department of Physics and Astronomy, University of
Kentucky, Lexington, Kentucky, 40506-0055}

\date{\today}

\begin{abstract}

The thermal conductivity $\kappa$ of the layered $s$-wave
superconductor NbSe$_2$ was measured down to $T_c/100$ throughout
the vortex state.  With increasing field, we identify two regimes:
one with localized states at fields very near $H_{c1}$ and one
with highly delocalized quasiparticle excitations at higher
fields. The two associated length scales are naturally explained
as multi-band superconductivity, with distinct small and large
superconducting gaps on different sheets of the Fermi surface.
This behavior is compared to that of the multi-band superconductor
MgB$_2$ and the conventional superconductor V$_3$Si.

\end{abstract}

\pacs{74.70.Ad, 74.25.Fy, 74.25.Qt, 74.25.Jb}

\maketitle


Multi-band superconductivity (MBSC) is the existence of a
superconducting gap of significantly different magnitude on
distinct parts (sheets) of the Fermi surface (FS). This unusual
phenomenon has recently emerged as a possible explanation for the
anomalous properties of some $s$-wave superconductors. Although
first experimentally observed over twenty years ago \cite{binnig},
the possible existence of MBSC has not often been considered since
then. The current interest in MBSC has been fueled by the peculiar
properties of the 40-K superconductor MgB$_2$, where the case for
MBSC is now rather compelling \cite{choi}. In particular, a gap
much smaller than the expected BCS gap has been resolved in
tunneling experiments \cite{eskildsen}. One consequence of such a
small gap is the ability to easily excite quasiparticles, which,
for example, can make the properties of this $s$-wave
superconductor similar to those of $d$-wave superconductors.

Based on angle-resolved photoemission (ARPES) measurements, it has
recently been proposed that the 7-K layered superconductor
NbSe$_2$ is also host to MBSC \cite{yokoya}. A sizable difference
in the magnitude of the superconducting gap  was found on two sets
of Fermi surface sheets, with no detectable gap on the smallest
sheet. However, these measurements were only performed at 5.3~K.
It is clearly of interest to shed further light on this
sheet-dependent superconductivity by performing bulk measurements
down to low temperatures.

In this Letter, we report a study of heat transport in NbSe$_2$
down to $T_c/100$ throughout the vortex state, providing further
evidence for MBSC. By measuring the degree of delocalization of
quasiparticle states in the vortex state, heat transport probes
the overlap between core states on adjacent vortices, {\it i.e.}
the size of the vortex core ($\sim \xi$), and hence the magnitude
of the gap ($\sim 1/\xi$). We resolve two regimes of behavior: one
limited to very low fields (up to $\sim 5 H_{c1}$), where
delocalization is slow and activated as in conventional
(single-gap) superconductors like V$_3$Si, and one for all other
fields up to $H_{c2}$ where quasiparticles transport heat
extremely well, as in unconventional superconductors with nodes in
the gap.


NbSe$_2$, a quasi-2D metal with hexagonal symmetry, displays a
transition to a charge density wave state around $T\simeq 35$~K.
Measurements of its FS at low temperatures, by both de~Haas-van
Alphen (dHvA) \cite{corcoran} and ARPES measurements
\cite{yokoya}, agree with band structure calculations that predict
a FS made of 4 or 5 sheets. These sheets divide into two groups: a
small $\Gamma$-centered pocket derived from the Se 4$p$ band
(denoted as $\Gamma$ band) and larger nearly two-dimensional
sheets derived from Nb 4$d$ bands. Scanning tunnelling
spectroscopy (STS) at 50 mK revealed a spectrum that is consistent
with a distribution of gaps that range from 0.7 to 1.4 meV
\cite{hess2}. In the vortex state at very low fields (near
$H_{c1}$), a zero-bias conductance peak, characteristic of states
localized in the vortex core, was observed at the vortex center
\cite{hess1}.


The thermal conductivity $\kappa$ of NbSe$_2$ was measured in a
dilution refrigerator using a standard technique \cite{boaknin}.
Measurements where made at temperatures increasing from $50$~mK
and in magnetic fields ranging from $0$ to $6$~T, applied parallel
to the $c$-axis and perpendicular to the in-plane heat current.
The sample was cooled in field to ensure field homogeneity.
Measurements as a function of field at fixed temperature resulted
in nearly no difference as compared to the field-cooled data (see
Fig.~2a).
\begin{figure}
\resizebox{\linewidth}{!}{\includegraphics{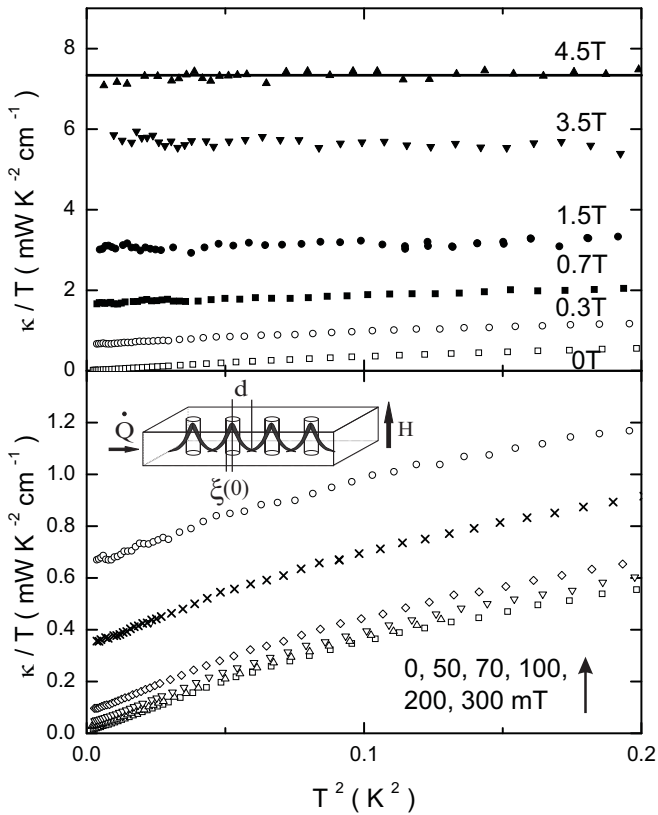}}
\caption{Thermal conductivity of NbSe$_2$ at several applied
fields, plotted as $\kappa/T$ vs $T^2$. The solid line indicates
the value expected from the Wiedemann-Franz law as obtained from
resistivity measurements at $H=4.5$~T. The field is applied
parallel to the $c$-axis and perpendicular to the heat current
$\dot{Q}$.}
\end{figure}
\noindent

The sample, a rectangular parallelepiped with dimensions 1.2
$\times$ 0.5~mm in the plane, and $0.1$~mm along the $c$-axis, is
from the same batch as the sample used by Sonier {\it et al.}
\cite{sonier1,sonier2}, and has a superconducting transition
temperature $T_c=7.0$~K with a width $\delta T_c = 0.1$~K. It was
cleaved to provide six fresh surfaces for silver paint contacts,
with resistances at low temperatures of roughly $20~$m$\Omega$.
The residual resistivity ratio is $40$ ($\rho_0\simeq 3 ~
\mu\Omega ~ cm$), and the upper and lower critical fields are
respectively $H_{c2}=4.5$~T and $H_{c1}=20$~mT for $H\parallel c$.
The coherence length estimated from $H_{c2}$ is
$\xi(0)=85~\rm{\AA}$.

The thermal conductivity of NbSe$_2$ is plotted in Fig.~1, as
$\kappa/T$ against $T^2$. This enables a separation of the
electronic and the phononic thermal conductivities, since the
asymptotic $T$ dependence of the former as $T \to 0$ is linear
while that of the latter is cubic. The electronic thermal
conductivity $\kappa_0/T$ is thus obtained as the extrapolated $T
\to 0$ value. In zero field, $\kappa_0/T = 0.000 \pm 0.005$~mW~K
$^{-2}$~cm$^{-1}$, a clear indication that NbSe$_2$ is an $s$-wave
superconductor with a fully gapped excitation spectrum. However,
by applying a small magnetic field ($H \geq H_{c1}$), an
electronic contribution develops as a rigid shift from the $H=0$
curve in Fig.~1. At higher fields ($H \geq 1.5~$T), the electronic
contribution dominates the conduction over the entire temperature
range and $\kappa/T$ is constant in temperature within our
experimental resolution. Above $H_{c2}$, the Wiedemann-Franz (WF)
law is satisfied and the thermal conductivity saturates.
\begin{figure}
\resizebox{\linewidth}{!}{\includegraphics{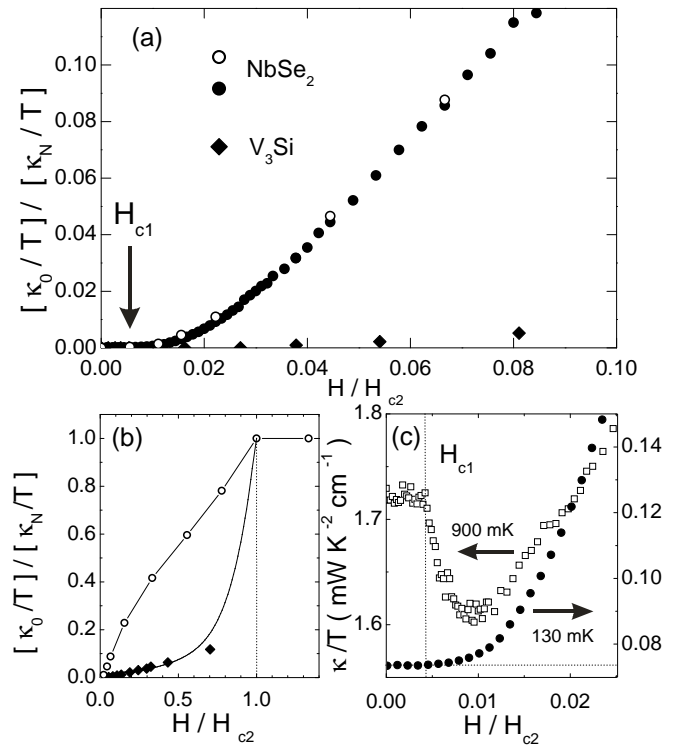}}
\caption{ (a,b) Thermal conductivity of NbSe$_2$ (empty circles)
and V$_3$Si (diamonds) at $T \to 0$ vs $H$, normalized to values
at $H_{c2}$. Filled circles come from a sweep in field at
$T=130$~mK from which the $H=0$ thermal conductivity (phononic
contribution) has been subtracted. The thick solid line in (b) is
a theoretical curve for the thermal conductivity of V$_3$Si
\protect\cite{dukan}. The thin line is a guide to the eye. (c)
$\kappa/T$ vs $H$ for NbSe$_2$ at $T=130$~mK (circles) and
$T=900$~mK (squares). The latter shows a typical drop in the
phononic thermal conductivity at $H_{c1}=20$~mT. The former shows
that the electronic thermal conductivity starts to increase right
at $H_{c1}$ but has a slow activated-like behavior for fields
below $0.03~H_{c2}$.}
\end{figure}
\noindent
\begin{figure*}
\resizebox{\linewidth}{!}{\includegraphics{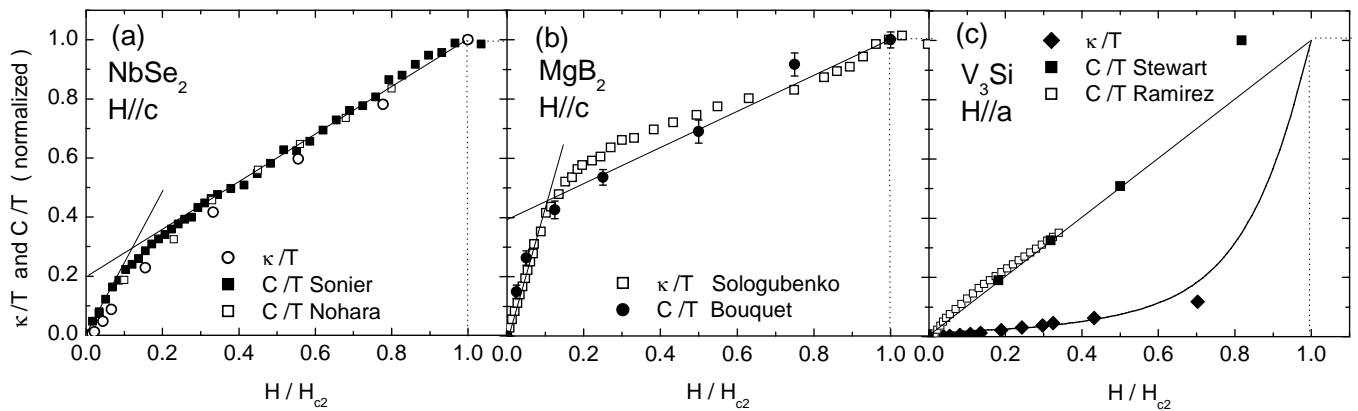}} \caption{
(a) Thermal conductivity and heat capacity of NbSe$_2$ normalized
to the normal state value vs $H/H_{c2}$. The heat capacity was
measured in two different ways: i) at $T=2.4$~K on the same
crystals as used in this study \protect\cite{sonier1}, and ii)
extrapolated to $T \to 0$ from various temperature sweeps on
different crystals \protect\cite{nohara1}. (b) Equivalent data for
MgB$_2$ single crystals \protect\cite{solo,bouquet}. (c)
Equivalent data for V$_3$Si, with a theoretical curve for
$\kappa/T$ \protect\cite{dukan}. The specific heat is measured at
$T=3.5$~K \protect\cite{ramirez} and extrapolated to $T=0$
\protect\cite{stewart}. The straight line is a linear fit. The
thermal conductivity is seen to follow the specific heat very
closely for both NbSe$_2$ and the multiband superconductor
MgB$_2$. It does not, however, for the conventional $s$-wave
superconductor V$_3$Si.}
\end{figure*}
\noindent

The $\kappa_0/T$ values are plotted as a function of $H$ on a
reduced scale in Fig.~2. Also plotted is a field sweep at
$T=130$~mK from which the zero-field value (the phononic
contribution) has been subtracted. As seen in Fig.~2c, heat
conduction starts to increase right at $H_{c1}$ in what could be
qualified as an activated behavior, although in a very limited
range of fields: $H_{c1} \leq H \leq 0.03 H_{c2}$. (The value of
$H_{c1}$ is determined in situ as the drop in the phonon $\kappa$
due to vortex scattering.) At higher fields, $\kappa_0/T$
increases rapidly, {\it i.e.} faster than ($H/H_{c2}$)$
~\kappa_N/T$, where $\kappa_N/T$ is the normal state value. This
shows the presence of {\it highly delocalized quasiparticle states
almost throughout the vortex state of NbSe$_2$}.

This is in stark contrast to the behavior expected of a type-II
$s$-wave superconductor. Indeed, when a field in excess of
$H_{c1}$ is applied, and vortices enter the sample, the
conventional picture is that the induced electronic states are
{\it localized} within the vortex cores. As one increases the
field, the intervortex spacing $d\simeq \sqrt{\Phi_0/B}$
decreases. The localized states in adjacent vortices will have an
increasing overlap leading to enhanced tunneling between vortices,
and the formation of conduction bands. Strictly speaking, the
electronic states are actually always delocalized but with
extremely flat bands at low fields \cite{yasui}. As these
gradually become more dispersive, the thermal conductivity should
increase accordingly and grow exponentially with the ratio
$d/\xi$, as is indeed observed in Nb \cite{vinen}.

A better point of comparison for NbSe$_2$ is V$_3$Si, an extreme
type-II superconductor with comparable superconducting parameters
($T_c=17$~K, $\xi=50~\rm{\AA}$). This is done in Fig.~2, where the
thermal conductivity of V$_3$Si is seen to grow much more slowly
with $H$ than that of NbSe$_2$, as described by theory
\cite{dukan}. Quantitatively, at $H=H_{c2}/20$, $\kappa_0/T =
\frac{1}{20} \times \kappa_N/T$ for NbSe$_2$ and $\frac{1}{400}
\times \kappa_N/T$ for V$_3$Si \cite{v3siWF}. Note that the
samples compared in Fig.~2 are in the same regime of purity. From
the standard relation $\xi(0)=0.74\xi_0 [\chi (0.88\xi_0
/l)]^{1/2}$, we obtain for V$_3$Si $\xi_0/l=0.13$, with
$\xi(0)=50~\rm{\AA}$ from $H_{c2}$ and $l=1500~\rm{\AA}$ from dHvA
\cite{janssen}. This is similar to the value of $0.15$ for
NbSe$_2$ \cite{takita}.

The high level of delocalization in NbSe$_2$ is a clear indication
of either a gap with nodes ({\it e.g.} $d$-wave) or a nodeless gap
which is either highly anisotropic or small on one FS and large on
another. A gap with nodes is ruled out by the absence of a
residual term in the thermal conductivity in zero field
\cite{graf}.

It is revealing to compare NbSe$_2$ to MgB$_2$, for which the
thermal conductivity has a similar field dependence. Strikingly,
$\kappa$ follows roughly the same field dependence as the specific
heat $C$ for both NbSe$_2$ and MgB$_2$. This is shown in Fig.~3
where $\kappa(H)$ and $C(H)$, are plotted on a reduced field scale
for single crystals of NbSe$_2$ \cite{sonier1,nohara1}, MgB$_2$
\cite{solo,bouquet}, and V$_3$Si \cite{ramirez,stewart}, with $H
|| c$ for hexagonal NbSe$_2$ and MgB$_2$, and $H || a$ for cubic
V$_3$Si. In conventional superconductors like V$_3$Si, $\kappa(H)$
and $C(H)$ are very different (see Fig.~3c) because the excited
electronic states are largely localized.

MgB$_2$ is a well established case of MBSC with a small gap on one
FS ($\Delta_{\pi}=1.8$~meV) and a large gap on the other
($\Delta_{\sigma}=6.8$~meV). The field dependence of its heat
capacity is well understood in this context \cite{bouquet}, with a
distinctive shoulder at a field of $H_{c2}/10$ (see Fig.~3b)
\cite{srruo}. A similar shoulder is also manifest in NbSe$_2$
around $H_{c2}/9$ (see Fig.~3a). Empirically, the striking fact
that heat transport and heat capacity have the same field
dependence in both materials points to a common explanation, and
hence suggests that {\it NbSe$_2$ is host to multi-band
superconductivity} \cite{kusunose}. This is consistent with recent
ARPES measurements at $T=0.8~T_c$ \cite{yokoya}.

\begin{figure}
\resizebox{\linewidth}{!}{\includegraphics{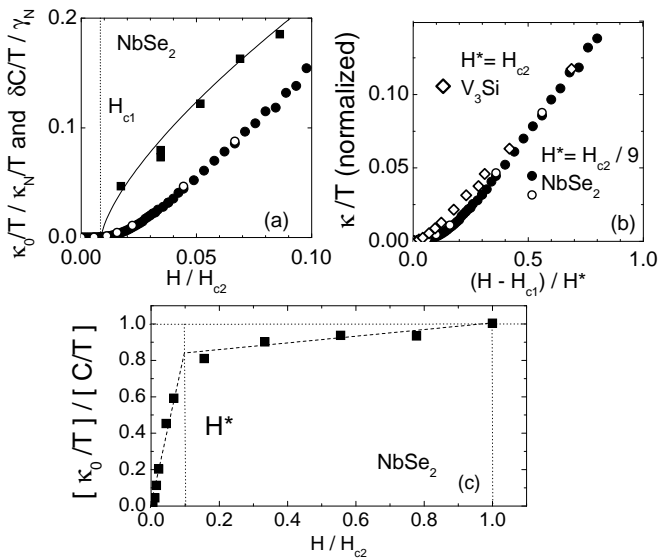}}
\caption{ (a) Thermal conductivity (circles) and the field
evolution of the heat capacity $\delta C/T = [C(T,H)-C(T,H=0)]/T$
(squares) \protect\cite{sonier1} of NbSe$_2$ normalized to the
normal state values vs $H/H_{c2}$ at very low fields. The line is
a guide to the eye. (b) Normalized thermal conductivity vs $(H -
H_{c1})/H^*$ for NbSe$_2$ and V$_3$Si with $H^*=H_{c2}/9$ and
$H_{c2}$ respectively. (c) Ratio of heat transport to heat
capacity in NbSe$_2$. }
\end{figure}
\noindent

In conventional superconductors, the delocalization of vortex core
bound states occurs gradually on the scale of $H_{c2}$, and the
characteristic length scale is $\xi(0) \simeq \sqrt{\Phi_0/2\pi
H_{c2}}$. It appears that in NbSe$_2$ (and MgB$_2$), there are two
characteristic length scales for delocalization: $\xi^*$ and
$\xi(0)$. To see this, we focus on the low field region. For
NbSe$_2$, both $\kappa/T$ and $C/T$ have been measured with high
precision on the same crystals, thereby making a detailed
comparison possible. Fig.~4a shows the comparison for fields below
$H_{c2}/10$, where the two do not coincide: $C/T$ increases
abruptly above $H_{c1}$ while $\kappa/T$ grows slowly, in an
activated way. This is consistent with the presence of localized
states at very low fields as imaged by STS \cite{hess2,hess1}.
Then this behavior gives way to a rapid increase of the thermal
conductivity at fields above $0.03~ H_{c2}$. This is a clear
indication that the field scale associated with delocalization in
NbSe$_2$ is much smaller than $H_{c2}$.

In fact, we can scale the behavior of the low-field thermal
conductivity of NbSe$_2$ to that of V$_3$Si using $H^*=H_{c2}/9$
(Fig.~4b). This is also seen clearly if we plot the ratio of the
thermal conductivity to the specific heat (Fig.~4c) which measures
the degree of delocalization. The ratio is seen to have two
regimes: a rapid increase below $H^* \simeq H_{c2}/10$ and a slow
one above. In summary, the second length scale associated with
delocalization in NbSe$_2$ is $\xi^* \simeq \xi(0) / \sqrt{9} =
\xi(0)/3$. This is consistent and may explain naturally the
shrinking of the vortex cores observed with muon spin rotation
\cite{sonier2}.

Considering the fact that the upper critical field is related to
the superconducting gap by $H_{c2} \propto \Delta^2 / v_F^2$ where
$v_F$ is the Fermi velocity, we estimate the gap to vary over the
FS by a factor of 3 ($\Delta^* \simeq \Delta_0/3$). (Note that we
assume $v_F$ to be constant, within the direction of the
$ab$-plane, as found by band structure calculations
\cite{corcoran}.) A variation of this order was reported earlier
from STS measurements \cite{hess2}, where the spread in $\Delta$
was measured to be between 0.7~meV and 1.4~meV. Moreover,
theoretical modelling of STS features associated with the vortex
cores \cite{hayashi} and efforts to model $C(T)$ \cite{kobayashi}
were found to require an anisotropy in the gap of a factor of 3
and 2.5, respectively.

Our results are consistent with dHvA measurements \cite{janssen}:
while extended quasiparticles are seen deep into the vortex state
in both NbSe$_2$ and V$_3$Si, the additional damping attributed to
the superconducting gap below $H_{c2}$ increases more slowly in
NbSe$_2$. Corcoran {\it et al.} measured the electron-phonon
constant $\lambda_{e-ph}=0.3$ for the $\Gamma$ band whereas they
extract an average value of $\lambda_{e-ph}=1.8$ from specific
heat \cite{corcoran}. The origin of MBSC may lie in the fact that
$\lambda_{e-ph}$ is smaller for the $\Gamma$ band. This suggests
that superconductivity originates in the Nb 4$d$ bands and is
induced onto the Se 4$p$ $\Gamma$ pocket.

In summary, measurements of heat transport in the vortex state of
NbSe$_2$ at low temperatures reveal the existence of highly
delocalized quasiparticles down to fields close to $H_{c1}$. This
is in striking contrast with what is expected in a $s$-wave
superconductor where well-separated vortices should support only
localized states, as is observed in V$_3$Si. We identify two
characteristic length scales that govern the destruction of
superconductivity in NbSe$_2$: the usual one associated with
$H_{c2}$ and another associated with a much smaller field $H^*
\simeq H_{c2} / 9$.  We attribute this to multi-band
superconductivity, whereby the gap on the pocket-like $\Gamma$
band is approximately 3 times smaller than the gap on the other
two Fermi surfaces.



We are grateful to H.-Y. Kee, Y.B. Kim, W.A. MacFarlane, K.
Machida, A.G. Lebed and K.B. Shamokhin for stimulating
discussions. We thank Sa\u{s}a Dukan for providing us with
theoretical calculations. This work was supported by the Canadian
Institute for Advanced Research and funded by NSERC of Canada.



\begin{references}
\bibitem{v3siWF} For V$_3$Si, $\kappa_N/T$ is taken to be $L_0/\rho_0$,
from the WF law where $L_0 = \pi ^2k_B^2/3e^2$ and $\rho_0$ is
extrapolated from the resistivity. We use $H_{c2}(0)=18.5~$T
\cite{janssen}.
\end{references}
\end{document}